\documentclass{aastex631}
\usepackage{amsmath, amssymb}
\usepackage{enumitem}

\usepackage{float}
\usepackage{comment}
\usepackage{graphicx}

\usepackage{indentfirst}

\begin{document}

\title{Seismic constraints on the spin evolution of slowly rotating young intermediate-mass stars}

\author[0009-0007-5514-4553 ]{Kunal H. Singh}
\affiliation{St. Xaviers College, Fort, Mumbai 400001, Maharashtra, India}

\author[0009-0009-1287-6521]{Subrata Kumar Panda}
\affiliation{Tata Institute of Fundamental Research, Colaba, Mumbai 400005, Maharashtra, India}

\author[0000-0003-2896-1471]{Shravan M. Hanasoge}
\affiliation{Tata Institute of Fundamental Research, Colaba, Mumbai 400005, Maharashtra, India}

\author[0000-0001-8699-3952]{Siddharth Dhanpal}
\affiliation{Tata Institute of Fundamental Research, Colaba, Mumbai 400005, Maharashtra, India}

\begin{abstract}
$\delta$ Scuti stars are hot, rapid rotators and are a poorly understood class of pulsators. Asteroseismology provides the only means with which to probe their interior dynamics. However, their complex and unexplained oscillation patterns restrict analyses to only a small fraction with interpretable pulsations. Here, we identify 5381 $\delta$ Scuti stars from 63 sectors of TESS observations, of which 300 had interpretable oscillations, with 24 showing rotational splittings. We inferred compositions and ages ($\tau$) for the 300 stars finding them in near-ZAMS states (\citealt{Bedding2020}), and measured the mean envelope rotation rates ($\langle f_{\rm rot} \rangle$) for 24 of them. Analyzing their age-dependent rotation, we found these stars essentially exhibit weak-to-no spindown, while evolving past the ZAMS across a narrow time-span during which they show regular pulsations. A quantitative fit to their spin-evolution results in a trend $f_{\rm rot} (d^{-1}) \propto (\tau/{\rm Gyr})^{-0.048\pm0.016}$, much slower than the spindown of cooler late-type stars due to magnetic braking (Skumanich’s law: $f_{\rm rot} (d^{-1}) \propto (\tau/{\rm Gyr})^{-0.5}$). Based on stellar evolution calculations, we show this weak spindown is consistent with the gradual increase in their moment-of-inertia. 
\end{abstract}

\section{Introduction}\label{section:introduction}
$\delta$ Scuti stars are intermediate-mass pulsating stars located near the main sequence and the lower part of the instability strip in the Hertzsprung-Russell (HR) diagram. They are hot stars, with effective temperatures ranging between 6000 K to 9000 K(\citealt{Uytterhoeven2011A&A...534A.125U}). Their pulsations, primarily driven by the $\kappa$-mechanism (\citealt{Chevalier1971A&A....14...24C}), are observed in multiple radial and non-radial acoustic (p) modes and lower radial-order gravity (g) modes (\citealt{Handler_2009_dscuti}, \citealt{2021FrASS...8...55G}). In contrast to the p-mode vibrations of stochastic oscillators (e.g., red-giant and solar-like stars), which are driven by turbulent near-surface convection, the acoustic pulsations in $\delta$ Scuti stars are coherently excited by the opacity-induced $\kappa$ mechanism in their Helium ionization zones within their stellar envelopes. $\delta$ Scuti stars oscillate with relatively high pulsating frequencies.
They are also very fast rotators, with surface rotational velocities reaching up to 200 - 300 km/s (\citealt{royer_rotational_velocity}). \\

The fast rotation of $\delta$ Scuti stars introduces unusual features such as the extension of the central mixed region (\citealt{2000ASPC..210....3Breger}) and complex pulsation patterns. 
On one hand, the process of deciphering these features grants us a deeper understanding of these stars and the fundamental principles governing them. However, on the other, the very rapid rotation responsible for these distinctive features, also presents challenges that impede the thorough study of these stars through asteroseismology. For non-rotating and slowly rotating stars, the pulsation modes may be represented using spherical quantum numbers ($n,\ell,m$). Associating these quantum numbers with a particular pulsating mode (`mode identification') is an essential step in constructing accurate models of structure and rotation.  
However, individual modes of fast rotators, which deform into ellipsoids, cannot be represented with unique sets of ($n,\ell,m$) since spherical harmonics are not the correct basis. Linear combinations of spherical quantum numbers can however be used to represent these modes, although this significantly complicates mode identification. We refer to \cite{Mirouh} for detailed insights into the identifications of pulsation modes in rapidly rotating stars. \\

The asymptotic theory of stellar oscillations (\citealt{2010aste.book.....A, Garc_a_2019}), of foundational importance in asteroseismology, states that in the regime $n \gg \ell$, p modes of the same degree ($\ell$) and successive radial orders ($n$) are equally spaced in frequency by a separation $\Delta \nu$. This helps in the easy and efficient identification of modes and also provides a foothold for realistic model creation. The pulsation frequencies determined by these models may then be compared with observed frequencies and aid in finding important stellar parameters such as age, mass, composition, etc. Hence, finding stars that display regular frequency spacings helps in shedding light on their internal structure and dynamics. However, $\delta$ Scuti stars pulsate in low radial-order overtones, i.e., $n$ and $\ell$ are comparable, and hence may not fall in the asymptotic regime.

\cite{Reese_2008} showed that for stars rotating at nearly 50\% of their Keplerian breakup velocity, `island modes', which are rotational counterparts of lower-degree modes, show regular frequency spacings in their power spectra, akin to asymptotic theory. Several studies have reported detections of regular frequency separations in $\delta$ Scuti oscillation spectra (\citealt{Matthews2007CoAst.150..333M}, \citealt{Garcia2009A&A...506...79G}, \citealt{2011MNRAS.414.1721Breger}, \citealt{2011_Zwintz},  \citealt{Paparo_2013}, \citealt{J.C_mean_density_2013}). A recent paper by \citet{Rieutord2024}, on the fast-rotating $\delta$ Scuti Altair, demonstrated the presence of regular frequency patterns in its oscillation spectrum, offering valuable insights into the modeling of such stars.
In an important paper in the field, \citet{Bedding2020} found 60 $\delta$ Scuti stars which exhibited regular frequency separations in their periodograms from the first 9 sectors of TESS observations. This work motivates our present search, which in addition to providing constraints on the internal structure, allows us to attempt the detection of systematic rotational splittings.

Limitations inherent in ground-based observations of stellar pulsations, such as atmospheric turbulence, light pollution, weather patterns, etc., diminish the quality and continuity of data collection.  In this context, the space-based missions CoRoT (\citealt{2009IAUS..253...71B}), MOST (\citealt{MOST2003PASP..115.1023W, MOST2004Natur.430...51M}), Kepler (\citealt{borucki2010Sci...327..977B}) and TESS (\citealt{Ricker_TESS}) have played a key role in providing us with high-quality, uninterrupted observations of millions of stars. TESS, a NASA mission, stands as an exemplar of this transformation by capturing high-precision photometric time series from the entire sky. This has in turn yielded a considerably larger dataset for asteroseismic studies compared to previous missions. TESS surveys the sky in small slices, termed sectors. Each sector is of dimension $24^\circ \times 96 ^\circ$  and it takes TESS approximately 27 days to scan each sector. In the primary mission, TESS surveyed each hemisphere in 13 sectors and subsequently, in their 27-month-long first-extended mission, they again covered both hemispheres over sectors ranging from 27 to 55. The offset in the field of view of the extended mission helps in closing gaps in the sky left during the primary mission. The second-extended mission will focus on the 56$^{\rm th}$ to 96$^{\rm th}$ sectors.
\par

As reported in \citealt{Bedding2020}, identifying regularly oscillating $\delta$ Scuti stars is challenging since it involves manually picking candidate stars followed by interactively inspecting their \'echelle diagrams through a gradual adjustment of $\Delta \nu$.
However, owing to the substantially larger data volume than that used by \citealt{Bedding2020}, extending this approach to stars across 63 sectors poses an algorithmic challenge. To accelerate this discovery process, we developed an `auto-correlation' method that provides us with a reliable estimate of pulsation regularity and the related $\Delta \nu$.

\section{Materials and Methods}
\subsection{Target selection}\label{section:target selection}
We obtained short-cadence (120 seconds) light-curves for the first 63 sectors of TESS from the\dataset[MAST ]{https://mast.stsci.edu/portal/Mashup/Clients/Mast/Portal.html} database as they are suitable for sampling high-frequency pulsations, such as in $\delta$ Scuti stars. The database comprises 15,000 to 20,000 FITS files of lightcurves captured at 2-minute cadence in each sector. 
To avoid false positives and misinterpretation of modes, we removed the external sources of periodicity and other object types by taking the steps described in Appendix A and were left with a sample of 292,684 stars.

\subsection{Characterising $\delta$ Scuti stars}\label{section:characterizing dscuti}

\begin{figure*}
    \centering
    \includegraphics[width=0.8\textwidth]{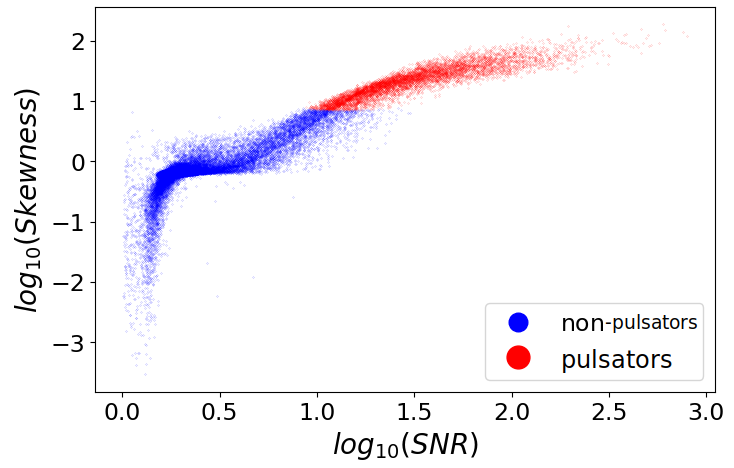}
    \caption{2D phase-space diagram where likely pulsators marked in red and non-pulsators in blue.}
    \label{fig:gmm}
\end{figure*}

FITS files of light-curves were read using the \texttt{lightkurve}  (\citealt{2018ascl.soft12013L}) Python package. We then rearranged the light-curves of the same object across different sectors in order of increasing sector number and then unified them using the ‘stitch’ function from the \texttt{lightkurve} package, thereby enhancing the quality of the signal. 

Following the criteria outlined in \citealt{2019MNRAS.485.2380M} and \citealt{Bedding2020}, we utilized the signal-to-noise ratio (SNR) and skewness to discern between $\delta$ and non-$\delta$ Scuti stars. 

An initial step involved the computation of the Lomb-Scargle periodograms (\citealt{Lomb1976}, \citealt{Scargle}) of the light-curves using \texttt{lightkurve}.
Along the lines of \citealt{Bedding2020}, we prefer selecting stars with high-frequency pulsations because they are more likely to be pure acoustic modes, and are expected to maintain equal frequency spacing, as dictated by the asymptotic theory. In the mid-frequency range, the spectrum is dense, exhibiting a mixed-mode nature due to the proximity of the gravity and acoustic cavities. The low frequency part of the spectrum supports buoyancy modes, and also shows long-period variability such as tidally induced modes, harmonics of rotation frequency, Rossby modes, harmonics of binary-revolution frequencies, motions of surface heterogeneities (e.g., spots), the orbital frequency of the observing satellite, etc. Gravity modes in rapidly rotating stars can have frequencies even beyond $5 {\rm d}^{-1}$.  In general, the higher-frequency end of the pulsation spectrum is considered to fully belong to the p-mode regime. Therefore, we discarded the lower-frequency portions of the power spectra and calculated the SNR and skewness using spectra in the range $30 - 95 \rm d^{-1}$, similar to \cite{Bedding2020}, in order to focus the analysis on pure acoustic modes in the asymptotic regime. The SNR for each star was calculated by dividing the power associated with the highest mode peak by the noise level, where noise corresponds to the $95^{\rm th}$ percentile of the entire amplitude spectra. Subsequently, we calculated the skewness of each spectrum using readily available routines from \textsc{scipy}, distinctly delineating the pulsation peaks from the surrounding background noise. 
All computations mentioned above were carried out in parallel on a 160-core computing node, making use of the \texttt{multiprocessing} (\citealt{multiprocessing_mckerns2012building}) Python package to effectively implement parallelization.

With these computed SNR and skewness values, we constructed a 2D phase-space plot (see Fig. \ref{fig:gmm}). This approach was motivated by Figure 7 of \cite{2019MNRAS.485.2380M}, where stars mainly segregate into two dense regions, comprising pulsators and non-pulsators. The stars in Figure \ref{fig:gmm} do not separate into distinct clusters. However, a closer look at the diagram indicates the presence of dense regions in the intervals $\log({\rm signal/noise}): (1.25-1.50)$ and $(0.2-0.5)$ and relatively lightly populated region in the range $\log({\rm signal/noise}) = (0.75-1.0)$. To separate the data points into two different groups, we exploited the Gaussian Mixture Models (\citealt {bishop2006gmm}) from \texttt{scikit-learn} (\citealt{scikit-learn}) package, which is an unsupervised method for performing density-based classification of data points. Without specifying signal-to-noise thresholds or skewness demarcations, we used the Gaussian Mixture Models to separate the data points into two groups, pulsators with high SNR and high skewness, and non-pulsators with low SNR and low skewness. Among the pulsators, we imposed the constraints $6000 \le T_{\rm eff} \le 9000$ K and $\log_{10}(L/L_\odot) > 0$ to characterize potential $\delta$-Scuti candidates. We visually inspected their spectra to ascertain whether they indeed exhibit typical $\delta$ Scuti-like pulsations. As a result, we were able to identify 5381 $\delta$ Scuti Stars from the first 63 sectors of TESS. It is possible that some of the non-pulsators classified here may be truly pulsators and However, the 5381 $\delta$ Scuti stars we identified here constitute a reliable and large catalogue of pulsators.

\subsection{Identifying regular pulsations and measurement of $\Delta\nu$}\label{section:regular pulsation method}

\begin{figure}
    \centering
    \includegraphics[width=0.49\textwidth]{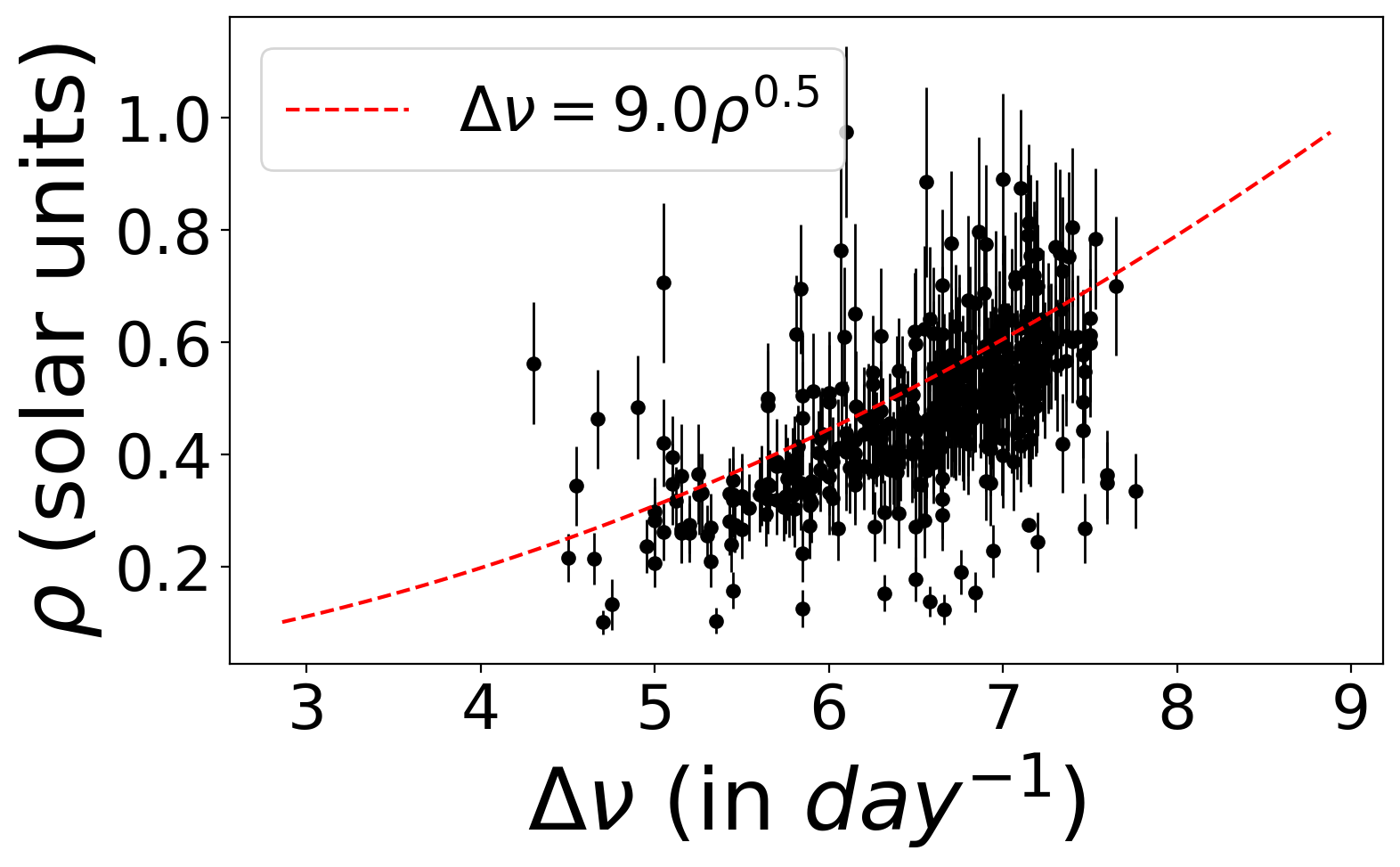}
    \caption{Variation of mean stellar density with respect to large frequency separation ($\Delta \nu$).}
    \label{fig:density}
\end{figure}

For a star exhibiting regular frequency pattern, the \'echelle diagram with an appropriate value of $\Delta \nu$ will show vertical ridges due to the alignment of modes of identical harmonic degree $\ell$ (Case A of figure \ref{fig:corr_graph comparison}). Such an \'echelle diagram may be imagined as a matrix with rows of length $\Delta \nu$ and vertical ridges as distinct columns with high-amplitude values (analogous to distinct degrees). 
Conversely for the same star, but with a $\Delta \nu$ far-off than the appropriate value, we see no such ridge-like structures in the \'echelle diagram (case B of figure \ref{fig:corr_graph comparison}).
Finally, for a star that does not has intrinsically regular pulsation pattern (case C of figure \ref{fig:corr_graph comparison}), the \'echelle diagrams will lack aligned ridges regardless of attempting various possible candidate values of $\Delta \nu$.

For a certain \'echelle diagram generated assuming a specific $\Delta \nu$, we calculate correlation coefficients between all successive rows and derive a mean of these coefficients, termed ‘mean correlation’. We notice that for case A, we will get a higher mean correlation value, while case B results in a lower mean correlation value, even though it is the same star albeit with the wrong $\Delta \nu$ value.
In contrast, for case C, the mean correlation values remain significantly lower across all the candidate $\Delta \nu$ values. 
Thus, we deduce that for a regularly pulsating star, if we calculate the mean correlation values for a range of possible candidate $\Delta \nu$ values, then the maxima of all the mean correlation values should correspond to the expected $\Delta \nu$ value.
We also find that the stars displaying greater regularity have higher global maxima values, essentially giving us a rough estimate on the extent of regularity of the star (refer supplementary figure \ref{fig:corr_graph comparison}). Because this method calculates the correlation coefficients by multiplying the amplitudes of successive \'echelle rows, it has the following limitation. Sufficiently small-amplitude regular-frequency patterns  may be obscured by high-amplitude modes and our present method cannot identify these subtle patterns. However, in this work, we have focused on identifying regular frequency patterns carrying high-amplitude modes, as in \citet{Bedding2020}, so that they can be verified through the use of \'echelle diagrams.

Utilizing this framework, we calculate the mean correlation values across a grid of possible $\Delta \nu$ values in the range 4 to 9 $d^{-1}$ with an interval of 0.002 $d^{-1}$ and find the maximum for each $\delta$ Scuti. This search range is an  extension of the typical $\Delta\nu$ values of $\delta$ Scuti stars ($5.3 - 7.7 {\rm d}^{-1}$), as reported in \cite{Bedding2020}.
Using the candidate $\Delta \nu$ value corresponding to the maximum, we construct \'echelle diagrams for these stars, as shown in figure \ref{fig:corr_graph comparison}. We then organize all the stars in descending order of their maximal values. This sorting approach optimizes our examination of regular stars, presenting the `highly' regular stars upfront and the `less' regular ones towards the end.

We calculated the uncertainties for $\Delta \nu$ by fitting a Lorentzian to the highest peak of the correlation function of each star. The center of the Lorentzian peak is deemed to be $\Delta \nu$ and half of the width is treated as the uncertainty. The Lorentzian function is defined according to equation \ref{eqn:Lorentzian },
\begin{equation} \label{eqn:Lorentzian }
\frac{A}{\pi}\left(\frac{\frac{\Gamma^2}{4}}{(x-x_0)^2+(\frac{\Gamma}{2})^2}\right),
\end{equation}
where A is the amplitude of the Lorentzian peak and $\Gamma$ is the Full Width Half Maximum. Although we used Lorentzians to fit the autocorrelation peaks, other profiles, e.g., Gaussian may also be adopted.
\par
Employing this method, we identified 300 $\delta$ Scuti stars displaying regular frequency separations from a total sample of 5381 $\delta$ Scuti stars. This approach significantly reduced the time and effort in this process.

We show the \'echelle diagrams of several $\delta$ Scuti stars in figure \ref{fig:regular_stars_echelle_diagram} that exhibit regular pulsations. Parts of the spectra appear clearer in the higher-frequency range, and turn fuzzier at the lower end, where various signals contribute to variability, as discussed in section 2.2. Multiple ridges are observed in some \'echelle diagrams, which may be attributed to signatures arising from higher-degree modes (e.g., $\ell = 2,3$), as suggested in \citealt{Bedding2020}.

Figure \ref{fig:density} demonstrates that the computed $\Delta \nu$ for the stars showing regular pulsation aligns with the standard scaling relation 
$\Delta \nu \propto \rho^{0.5}$ (\citealt{Ulrich1986ApJ...306L..37U}), where mean stellar density($\rho$) is obtained from TESS Input Catalogue hosted at the vizier database (\citealt{vizier2000A&AS..143...23O}). This relation has been theoretically predicted and verified in the literature (\citealt{J.C_mean_density_2013, García_Hernández_2015, García_Hernández2017, Rodríguez-Martín2020}), where the authors found that the form $\rho \sim \Delta\nu^\alpha$ fits the observations well, with $\alpha\approx$2.0. Recently, \cite{Bedding2020} also demonstrated $\Delta\nu \propto \rho^{0.5}$ for the $\delta$ Scuti stars in which they found regular frequency patterns. This correlation arises because $\Delta\nu$ is determined from modes in the quasi-asymptotic regime. The correlation is obscured by large scatter, which may be due to the direct effect of rotation (\citealt{cepher}) on mean density or indirect effect of gravity darkening (\citealt{grav_drk_scatter}) on photospheric properties used to evaluate the stellar density. The large-frequency separations ($\Delta\nu$) of the 300 stars show an asymmetric distribution (figure \ref{fig:dnu_histogram}) skewed towards a higher value, representing an enhanced population of relatively young stars in our samples, as $\Delta\nu$ is known to gradually decrease with stellar evolution. This implication is in agreement with the observed over-abundance of young stars, shown further in figure \ref{fig:modelfit}.

By identifying the $\Delta\nu$ of regularly oscillating $\delta$ Scuti stars, our study can effectively constrain their mean density.
We find that 6\% of all the $\delta$ Scuti stars discovered show regular frequency separations in their oscillation spectra, which is consistent with the findings reported in \citealt{Bedding2020}. 
This analysis can help us gain a preliminary understanding of the $\delta$ Scuti stars' evolution. Nevertheless, a sample of $\delta$ Scuti stars covering diverse evolutionary stages is imperative to investigate their detailed functioning (\citealt{whatCoRoT}). The question of why only 6\% of $\delta$ Scuti stars pulsate in a regular manner remains. All the 5381 $\delta$ Scuti stars in our sample and the 300 that show equally spaced acoustic mode patterns are homogeneously distributed across the sky (figure \ref{fig:skymap1}), which rules out specific regions of the galaxy preferentially harboring these stars. However, most of the 300 regular pulsators lie tightly clustered around the ZAMS line in the HR diagram (figure \ref{fig:hrd1}). This suggests that a majority of stars showing regular pulsation patterns are near-ZAMS stars, as also argued by \cite{Bedding2020}. Since such young stars possess pulsation cavities that are simpler in structure (\citealt{Handler_2009_dscuti}), they may be vibrating with purely acoustic modes, thus resulting in a cleaner spectra with equally spaced frequencies. With further main-sequence evolution (\citealt{Handler_2009_dscuti}), mixed modes emerge in the spectra, along with the development of avoided crossings, and mode amplitudes become more unpredictable -- which collectively complicate the detection of regular frequency patterns in evolved stars. In figure \ref{fig:hrd1}, a handful of possibly evolved stars with regular frequency patterns appear beyond the ZAMS line.

\subsection{Fitting stellar models}\label{section: fitting stellar model}
We attempt to infer the internal structure of these stars in an effort to appreciate features that drive their regular oscillations. Deploying a modified version of the conventional $\chi^2$ minimization between the spectroscopic-asteroseismic observations and the stellar-model grid available from \cite{murphy2023grid}, we intended to obtain the best-fit models ($M, Z, \tau$) for each of the 300 stars identified in Section \ref{section:regular pulsation method}. The model grid (\citealt{murphy2023grid}) comprises over 200,000 stellar models covering a wide range over mass, metallicity, and age. This grid also provides pulsation frequencies of radial and dipole zonal modes with frequencies up to $95 {\rm d^{-1}}$. For easier comparison against real spectra, we developed power spectra by constructing sinc profiles (\citealt{Dupret, PL2019, revPL2022}) of identical amplitudes around each of these eigenfrequencies. Although it does not exactly model the amplitude structure in the observations, it serves the purpose of frequency-profile matching between models and observations. The amplitude scale is not a concern because we normalize both spectra to allow for direct comparison.

Although rotation has not been implemented in this stellar model grid, it has been used by various authors (\citealt{m2020, m_precise_age, m_pl, scutt2023asteroseismology}) to model $\delta$ Scuti stars rotating at slow-to-intermediate rates. The grid incorporates convective overshooting using an exponentially decaying mixing scheme (\citealt{murphy2023grid}) implemented at the boundaries of convective cores and at the surface layers.

In several contexts, many authors (\citealt{Reeth2015, Bazot,  m_precise_age, Steindl}) have used $\chi^2$ minimization techniques to fit relevant parameters to astrophysical observations, including multiple asteroseismic oscillation frequencies. In this method, $\chi^2$ denotes the difference between model and observation.
The minimum-$\chi^2$ model ($\chi^2_{\rm min}$) is accepted as the best-fit solution and the 1-$\sigma$ uncertainty region is determined by measuring the spread in parameters of the models whose $\chi^2$ is less than $\chi^2_{\rm min} + \Delta$, where $\Delta$ is an offset dependent on the number of free parameters used in the fitting. In the present case, which involves three free parameters $(M,Z,\tau)$, $\Delta = 3.5$ (\citealt[Table 1]{Avni1976}). The asteroseismic term in the expression $\chi^2$ compares the oscillation frequencies of observations and the model possessing the same mode characteristics, thereby making prior mode identification a necessary requirement. This limits the applicability of this strategy to only a few stars. Thus, manual investigation of mode identities makes it difficult to automate this method to address a larger sample of stars. Left with the challenge of modeling 300 stars lacking proper mode labels, we resorted to adopting a modified version of $\chi^2$ minimization as we elaborate below.

Our merit function involves model-observation comparisons over two categories of terms, first, the comparison of $(L, ~T_{\rm eff}, ~\Delta\nu)$ and second, the similarity between the modeled and the observed spectra. Different stars are observed for different temporal durations ($T_{\rm obs}$) and therefore the spectral resolution ($1/T_{\rm obs}$) varies across the sample. For a minimum one sector of observation, TESS data allows a resolution of $0.037 ~{\rm d}^{-1}$. However, we need a common frequency resolution for all the stellar spectra in our sample since they need to be compared with thousands of model spectra having identical resolution. Thus, after computing the individual power spectra, we linearly interpolated their amplitudes on to a frequency grid spanning $1-95 ~{\rm d}^{-1}$ comprising $10000$ array elements (resolution of $\sim0.0094 ~{\rm d}^{-1}$). Due to this, we tend to over-sample stellar spectra which have less than 4 sectors of TESS data, and under-sample those which have been observed for longer. Choosing a frequency grid of higher resolution will increase computational cost without significantly impacting the inference. However, adopting a coarser resolution may result in important pulsation modes being neglected, leading to mis-identifications of best-fit solutions. Given a stellar observation characterized by ($L_\star, ~T_{\rm eff\star}, ~\Delta\nu_\star, ~{\rm spectra}_\star$), the stellar model represented by ($L_{\rm model}, T_{\rm eff; model}, ~\Delta\nu_{\rm model}, {\rm spectra}_{\rm model}$) is compared according to the following merit function
\begin{equation*}
    \chi^2_{\star, \rm model} = (L_\star-L_{\rm model})^2 + (T_\star-T_{\rm model})^2 + (\Delta\nu_\star-\Delta\nu_{\rm model})^2 + \left( \dfrac{1}{\rm spectra_\star \otimes spectra_{model}} \right)^2,
\end{equation*}
where `$\otimes$' represents the summed-correlation (or equivalently the `dot product'), which is a scalar value obtained by adding the element-wise products of two arrays. The inverse form of the last term was chosen considering that the $\chi^2$ metric may be reduced by minimizing the first three terms (differences in $L, T, \Delta\nu$), while at the same time maximizing the last term (spectral correlation). To avoid the dot product yielding very large values in the last term, we normalized each spectra according to ${\rm spectra_\star \to   spectra_\star / \sqrt{spectra_\star \otimes spectra_\star}}$ and ${\rm spectra_{model} \to spectra_{model} / \sqrt{spectra_{model} \otimes spectra_{model}}}$. Hence, the correlation term ${\rm spectra_\star \otimes spectra_{model}}$ is constrained so as to never exceed unity. The model with the lowest $\chi^2_{\star, \rm model}$ value may be considered the best fit, which happens when ($L_{\rm model}, T_{\rm eff; model}, ~\Delta\nu_{\rm model}$) are closest to ($L_\star, ~T_{\rm eff\star}, ~\Delta\nu_\star$) and the correlation term ${\rm spectra_\star \otimes spectra_{model}}$ is closest to unity.

To find the best model across the entire grid for a given star, one would need to compute the $\chi^2_{\star,{\rm model}}$ for all model samples and subsequently identify the set of parameters for which this assumes the lowest value. The four individual terms in the $\chi^2_{\star, \rm model}$ can vary over distinct scales (e.g., $\sim 10^{-2} L^2_\odot, ~\sim 10^4 K^2, ~\sim 10^{-2} {\rm d}^{-2}$), with $T_{\rm eff}$ dominating the other components. The appropriate weighting of each of these terms is important to generating accurate best-fit models. We could have scaled the first three difference terms by their $1 \sigma$ uncertainties but the final correlation term does not show well defined uncertainty. Trial and error led us to better solutions when each term is separately normalized by its corresponding largest value across the entire model grid. This ensures that each term in the $\chi^2_{\star, {\rm model}}$ stays between 0 and 1 regardless of the model and star. Equation \ref{eq:chisq} expresses these ideas in a mathematical form, with the symbol `$\mathcal{N}$' representing normalization across the entire grid.

\begin{align} \label{eq:chisq}
    \chi^2_{\star, \rm model} = 
    & \quad \mathcal{N} \left\{ (L_{\rm \star}-L_{\rm model})^2 \right\}
    + \mathcal{N} \left\{ (T_{\rm eff; \star}-T_{\rm eff; model})^2 \right\}
    + \mathcal{N} \left\{ (\Delta\nu_{\rm \star}-\Delta\nu_{\rm model})^2 \right\} \nonumber \\
    & + \mathcal{N} \left\{ ({\rm spectra_{\star} \otimes spectra_{model}})^{-2} \right\}
\end{align} 
This may be more clearly seen  when the  equation \ref{eq:chisq} is rephrased as

\begin{align*}
    \chi^2_{\star, \mathrm{model}} = 
    & \quad \frac{(L_{\star}-L_{\mathrm{model}})^2}{\genfrac{}{}{0pt}{}{\mathrm{max}}{\text{all models}} \left[ (L_{\star}-L_{\mathrm{model}})^2 \right]}  
    + \frac{(T_{\mathrm{eff; \star}}-T_{\mathrm{eff; model}})^2}{\genfrac{}{}{0pt}{}{\mathrm{max}}{\text{all models}} \left[ (T_{\mathrm{eff; \star}}-T_{\mathrm{eff; model}})^2 \right]} \\
    & + \frac{(\Delta\nu_{\star}-\Delta\nu_{\mathrm{model}})^2}{\genfrac{}{}{0pt}{}{\mathrm{max}}{\text{all models}} \left[ (\Delta\nu_{\star}-\Delta\nu_{\mathrm{model}})^2 \right]} \\
    & + \frac{(\mathrm{spectra}_{\star} \otimes \mathrm{spectra}_{\mathrm{model}})^{-2}}{\genfrac{}{}{0pt}{}{\mathrm{max}}{\text{all models}} \left[ (\mathrm{spectra}_{\star} \otimes \mathrm{spectra}_{\mathrm{model}})^{-2} \right]} 
\end{align*}

We prefer the inclusion of the spectral-correlation term in the $\chi^2$ metric over the commonly used eigenfrequency-comparison, especially because of ambiguities in assigning labels to observed oscillation modes. A similar spectral correlation technique was also used by \cite{Li_2022} for parameter-fitting to Kepler RGB stars.

With this setup, the best-fit model will have a very small $\chi^2_{\star, \rm model}$ and the model that maximally differs from observation will possess a $\chi^2_{\star, \rm model}$ around 4. For a particular star, the $\chi^2_{\star, \rm model}$ values obtained for all models leads to a Gaussian dip in the vicinity of the global minimum (similar to Fig. 9 of \citealt{m_precise_age} in the 3D phase space of ($M,Z,\tau$)). We computed the best-fit solutions by identifying the model associated with the lowest merit function, and obtained corresponding uncertainties by calculating the $1-\sigma$ parameter spread of the neighboring models. While calculating the expected values and uncertainties, we did not disentangle the three structure parameters, rather treating them as a single entity to retain their mutual correlations intact.
Equation \ref{eq:chisq} homogeneously weighs all four constraints to define the merit function for each model. Visually inspecting the match between the best-fit model's eigenmodes and physical properties with the corresponding observables, we noticed that equation~\ref{eq:chisq} provides reasonable fits to most stars. The result of the fitting algorithm is demonstrated for all 300 stars in the supplementary file3. It is important to keep in mind that our results do not represent the absolute best fits for the corresponding stars -- these inferences are subject to better estimations as the stellar evolution models improve with the inclusion of more realistic ingredients.

Finally, rotational mixing within the radiative envelopes, which is not incorporated in stellar evolution models, might play a significant role in perturbing the structure and pulsations of intermediate-mass main-sequence stars (\citealt{Pedersen_2022}). Hence, our reported best models based on $\chi^2_{\rm min}$ may not accurately capture all relevant aspects of stellar structure. Within our present capabilities, they offer the best potential representations to the overall observations of these targets.

\begin{figure*}
    \centering
    \includegraphics[width=\textwidth]{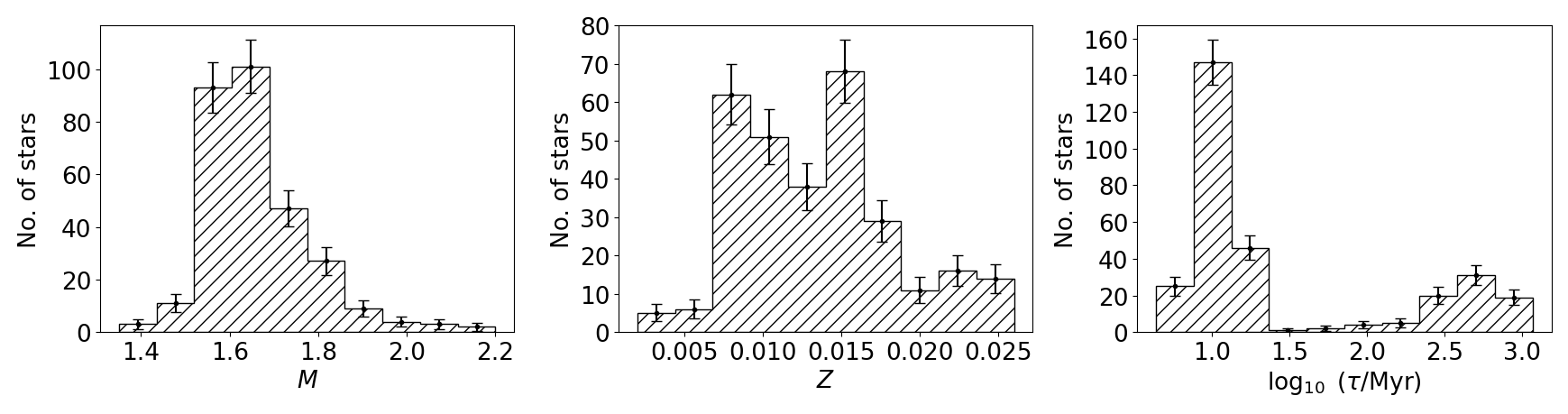}
    \caption{Distributions of stellar parameters $M$, $Z$, and $\log (\tau/{\rm Myr})$, obtained from fitting models to spectroscopic and seismic observations of 300 regular oscillators.}
    \label{fig:modelfit}
\end{figure*}

Figure \ref{fig:modelfit} indicates that, while mass and metallicity tend to cluster around $1.6 M_\odot$ and 1\% (or [Fe/H] $\sim -0.146$) respectively, the age distribution is bimodal. Most stars appear to have low metallicity as compared to the Sun, i.e., $Z_\odot = 0.019$. The [Fe/H] values for these stars as obtained from Gaia DR3 (\citealt{gaia_release}) also support this assertion (figure \ref{fig:Zcompare}).
Bimodality in age may arise from the fact that $\delta$ Scuti-like oscillations are often excited across a diverse evolutionary phases -- ranging from pre-main sequence to the terminal age main sequence, and occasionally even beyond. $\delta$ Scuti stars can even exhibit comparable asteroseismic properties (e.g., $\Delta\nu$ or individual mode frequencies) at very different evolutionary phases, which may cause degeneracy in their age determination, especially because we only have access to a limited number of observables. The left peak in the age-distribution figure may be pointing to the pre-main phase stars and the right, to those in the main sequence proper. However, most stars were found to be younger than 30 Myr, consistent with \cite{Bedding2020}, suggesting that stars with regular patterns are likely in near-ZAMS phases.

These inferences also possess the potential to help in characterizing open clusters in which $\delta$ Scuti stars may be present (\citealt{cluster_ds_Praesepe, cluster_ds_2022, cluster_ds_2023}), allowing for constraining the related parameters (\citealt{Handler_2009_dscuti}). 
Recently, \cite{2023ApJ...946L..10B_pleiades} identified the presence of several $\delta$ Scuti stars in TESS observations of the Pleiades cluster, underscoring a promising avenue for future investigations.

\section{Results and Discussion}

Stellar rotation is very important in inferring age and magnetism and plays a crucial role in influencing stellar structure (\citealt{10.1093/mnras/stu728_rotation_imp_magnet}).  Asteroseismology has been highly successful in measuring the internal rotation rates of various pulsating stars. Because of the number of orderly asymptotic modes with accurately labelled rotational splittings, rotation inferences of red giants (\citealt{Deheuvels2012}) and solar-like oscillators (\citealt{Benomar2015}) has met with success. Asteroseismologists are able to place constraints on their radial and latitudinal differential rotation rates (\citealt{Bazot, Benomar_2018}). Owing to their gravity-mode pulsations, whose mode identifications have recently become possible with high-resolution spectra from Kepler, notable progress has been made in inferring the rotation rates of rapidly spinning hot main-sequence stars, which include intermediate-mass $\gamma$ Doradus (\citealt{Reeth2015}), high-mass slowly pulsating B stars (\citealt{Pedersen2022}) and $\beta$ Cepheids (\citealt{bCep_rot_2004, bCep_rot_2008, Suárez_2009, bCep_rot_review}). This has generated several new insights, including cores rotating much faster than envelopes (\citealt{Deheuvels2012}), envelopes rotating slightly faster than cores (\citealt{df-rot-KIC11145123}), and cores slowing down with stellar age (\citealt{Pedersen2022}).

However, the identification of rotational multiplets in $\delta$ Scuti stars has long remained a formidable problem. Obstacles to this goal are (i) intrinsic amplitude suppression of certain multiplets, (ii) lower inclination angles, favouring zonal modes, or (iii) large (${\rm f_{rot} \sim \Delta\nu}$) asymmetric splitting of multiplets, disrupting traditional structure in spectra. While the first scenario has been explicitly observed in $\gamma$ Doradus stars as prograde mode excitation (\citealt{611Doradus}), the second may be justifiable for $\delta$ Scuti stars showing regular patterns, and the last point seems applicable to rapid rotators. Some of the success down this line includes, but is not limited to, the discovery of a dipole triplet in V 624 $\tau$ (\citealt{rotsplit-624tau}), spectroscopic identification of rotational splits in FG Vir (\citealt{FGVir-rotsplit}), observation of rotational multiplets in 31 CoRoT $\delta$ Scuti stars (\citealt{UNEXPECTED_SERIES}), and detection of four dipole doublets in HD 23642 (\citealt{rotsplit4doublets}) among others. However, simultaneous measurements of rotation and age on a larger sample are still lacking - which we are attempting to carry out in this work.  Our present sample is limited to slow rotators in a short near-ZAMS phase of evolution, and hence is not representative of the entire class of $\delta$ Scuti stars.

\subsection{Discerning rotational splittings in $\delta$ Scuti stars} \label{section:rotational splitting}

We investigated rotational splitting patterns exhibited by (dominantly) dipole modes in a selected group of 24 stars. This set was manually chosen by identifying spectra in which multiplet-like structure was seen across at least two radial orders of dipole modes. We show \'echelle diagrams of two different stars in figure \ref{fig:rotational_splitting_good}, where rotational splittings are observed across several radial orders, thus supporting the idea that they are multiplets arising from rotational perturbations.
The doublet patterns seen in the \'echelle diagrams are presumed to be $m=-1$ and $m=+1$ modes -- which, invoking traditional seismic interpretation, offer an avenue to infer the rotation rate.

Rotational splittings in $\delta$ Scuti stars can take on unexpected and complicated forms. For instance, it has been argued that prograde ($m=+1$) modes are selectively driven (\citealt{Reeth2015}) in $\gamma$ Doradus stars. Similarly, $\delta$ Scuti stars also show peculiar rotational splittings, e.g., \cite{rotsplit4doublets} has observed dipole splittings at radial order $n=1, 2, 3, 6$ with $m=0$ modes consistently absent (see their Figure 4 and 3) in all harmonics in the binary star HD 23642 (only $m=\pm 1$). Based on this, they conclude that the star is seen nearly edge-on. They determined the rotation frequency to be half the doublet spacing, which is also consistent with the synchronized binary orbital period of this system. In the 24 stars we examined, we noticed identical dipole doublet features in the \'echelle diagrams of 15 stars, with the exception of $m=0$ modes occasionally showing up at certain radial orders in  the remaining 9 stars. Therefore, along similar lines of argument, we assume that wherever we observe doublets, they may be identified as $m=\pm 1$.

Stars are likely to not rotate as rigid bodies. They exhibit radially as well as latitudinally varying rates of rotation, although the extent differs among different categories of stars. The rotational splittings observed in oscillation spectra are a consequence of the sum of internal rotation profiles weighted by the mode sensitivity. Only in the case of uniformly rotating stars is the mode splitting a direct measure of the rotation rate of the star. Although differential rotation has been well constrained in solar-like stars (\citealt{Benomar_2018}), little is known about this phenomenon in $\delta$ Scuti stars. \cite{df-rot-KIC11145123} estimated that main-sequence $\delta$ Scuti KIC 11145123 exhibited rigid rotation, with the envelope spinning only 3\% faster than the core. \cite{dfrotsolid} reported KIC 10080943 as an eccentric binary system with both the primary and secondary components comprising morphologically similar $\delta$ Scuti - $\gamma$ Doradus hybrids, in which they observed the envelope to rotate 1\% slower (3\% faster) than the core in the primary (secondary) component. \cite{Murphy2016} concluded that main-sequence star KIC 7661054 rotates approximately uniformly, with its surface rotating slightly faster than its core. Based on these lines of argument, we invoke the assumption of rigid body rotation for the stars we analyzed as a first approximation. Since we use the splittings of low-radial order p-modes (confirmed by matching with the best-fit model's pulsation), whose sensitivities peak in the outer layers, our measurements are likely indicative of near-surface (or outer envelope) rotation. Even in the presence of differential rotation within the outer stellar layers, our reported values may be considered to be the average envelope rotation. 

When accounting for only the effect of Corilois force and ignoring centrifugal deformation for a differentially rotating star, the frequencies of rotationally perturbed oscillation modes in the observer's frame may be written as (\citealt{df-rot-KIC11145123})
$$\nu_{\rm inertial} = \nu_{\rm co-rotating} + m(1-C_L) \int_0^R K_{n,\ell} (r) f_{\rm rot}(r) dr,$$
where $$C_L = \dfrac{\int_0^R \xi_h(2\xi_r+\xi_h)r^2\rho dr}{\int_0^R [\xi_r^2+\ell(\ell+1)\xi_h^2]r^2\rho dr}$$ is the Ledoux constant (\citealt{Ledoux}), and $$K_{n,\ell} = \dfrac{[\xi_r^2 + \ell(\ell+1)\xi_h^2 - 2\xi_r\xi_h - \xi_h^2]\rho r^2}{\int_0^R [\xi_r^2 + \ell(\ell+1)\xi_h^2 - 2\xi_r\xi_h - \xi_h^2]\rho r^2 dr}$$ is a weight factor known as the rotation kernel, which captures mode sensitivity as a function of stellar radius. $\xi_h$ and $\xi_r$ denote the horizontal and radial displacements of the mode eigenfunction, computed with respect to the equilibrium configuration.

The rotation kernels for p-modes are more sensitive to the outer stellar layers (Fig. 12 of \citealt{df-rot-KIC11145123}). With increasing radial order, their probing power shifts ever more towards the surface. Hence, denoting the average near-surface envelope rotation as $\langle f_{\rm rot}\rangle$, rotationally split frequencies may be written as
$$\nu_{\rm in.} = \nu_{\rm co.} + m(1-C_L) \langle f_{\rm rot}\rangle.$$
At moderate-to-fast rotation (\citealt{Lignières2006, Reese2006, Ballot2013}), the above equation representing first-order perturbation becomes inadequate to explain the  frequency splittings of $\delta$ Scuti stars. The effect of centrifugal deformation significantly perturbs different $m$ components, even including the $m=0$, thus causing the frequency spacing to become asymmetric. This is why the $m=0$ component does not appear for some spectra at the midpoint between the $m=\pm 1$ doublet. Combining both first- and second-order effects, the frequencies as measured in the inertial frame may be written as
$$\nu_{ n, \ell,m} = \nu_{ n, \ell} + m(1-C_L) \left<f_{\rm rot}\right> + (D_0 + m^2 D_1) \dfrac{\left<f_{\rm rot}\right>^2}{\nu_{ n, \ell}},$$
where the coefficients $\{D_0, D_1\}$ represent complicated functions of stellar structure and displacement eigenfunctions (\citealt{Suárez2006}). Retaining the second-order terms, the above equation may be used to express the average envelope rotation $\langle f_{\rm rot}\rangle$ in a form
$$\nu_{ n, 1, 1} - \nu_{ n, 1, -1} = 2(1-C_L) \langle f_{\rm rot}\rangle,$$
where the second-order effect appears redundant. This may be used to calculate the rotation rates from the sectoral dipole multiplets ($m=\pm 1$) without parameterizing the splitting asymmetry or the $m=0$ frequencies
\begin{equation} \label{eq:frot}
    \langle f_{\rm rot}\rangle = \dfrac{\nu_{m=+1} - \nu_{m=-1}}{2(1-C_L)}.
\end{equation}
Similar formulae have been theoretically and numerically (\citealt{Reese2009, Deupree_2011}) proposed to derive the rotation rates. This way of measuring the rotation frequency is reliable under the assumption that expanding up to the second-order effect is sufficient to explain the mode frequencies. We used the equation \ref{eq:frot} to calculate the average envelope rotation exploiting the dipole doublets seen in the 24 stars. For each star, we calculated the Ledoux constant from the best-fit model's appropriate oscillation modes that match the identified splitting in the observed spectrum. The Ledoux constant for p-modes are often very small, e.g., 0.03 for the stellar model of KIC 11145123 (\citealt{df-rot-KIC11145123}), and ranges from 0.007 to 0.022 for different modes in various stars that we studied. Because $1-C_L \approx 1$, the inferred rotation rates are not highly sensitive to the Ledoux constant.

\begin{figure*}
    \centering
    \includegraphics[width=\textwidth]{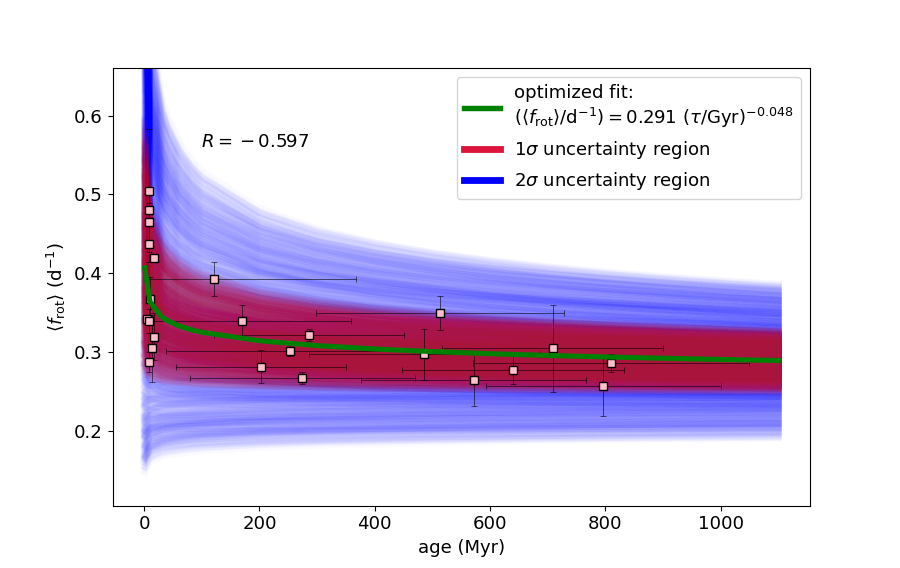}
    \caption{ Evolution of average envelope rotation with age of slow-rotating near-ZAMS $\delta$ Scuti stars. These stars, {\it covering a tiny phase space of the rotation-age landscape}, show weak-to-no spin down. The Pearson correlation coefficient for the dependence between $\langle f_{\rm rot} \rangle$ and age yields $R = -0.597$. Similar analysis results in a correlation of $R = -0.656$ between $\log\tau$ and $\log f_{\rm rot}$. Larger rotation spread for the youngest stars further reduces the expected correlation.    The green line shows the best-fit curve to the data,    which appears to follow a $\langle f_{\rm rot}\rangle \propto (\tau/{\rm Gyr})^{-0.048\pm0.016}$ trend, with $f_{\rm rot}$ measured in $\rm d^{-1}$.  The 1- and 2-$\sigma$ uncertainty regions are shown as solid patches by plotting the characteristic functions with parameters sampled within the $1\sigma$ or $2\sigma$ uncertainty ellipses centered at the best fit values.
    }
    \label{fig:skumanich}
\end{figure*}

\subsection{Gyrochronological Implications}

Late-type stars generate a strong magnetic field due to the motion of the ions in their convective zone, which acts as a dynamo. With increasing age, stellar angular momentum decreases as the magnetic field causes ionized wind to blow outward, in turn leading to a decrease in their rotation rate - termed magnetic braking. Stars with higher rotation rates have stronger magnetic fields (\citealt{2007AAS...21110327S}), thereby leading to a more rapid spindown. This feedback mechanism leads to the convergence of rotation rates over time, and they eventually follow the Skumanich spin down law, $f_{\rm rot} \propto t^{-0.5}$ (\citealt{skumanich1972ApJ...171..565S}). This phenomenon has been consistently studied in open clusters of ages 100-650 Myr, Hyades (650 Myr, \citealt{Radick1987ApJ...321..459R}), Pleiades (500 Myr, \citealt{Barnes2007ApJ...669.1167B}), Coma Berenices(650 Myr, \citealt{collier}), M37 (550 Myr, \citealt{Hartman_2009}). This limitation is a big hurdle as clusters of age $\sim$ 500 Myr are quite rare. The spindown relation is also useful in inferring ages for field stars as shown by various studies (\citealt{Mamajek2008ApJ...687.1264M}, \citealt{Meibom_2009}, \citealt{McQuillan2014ApJS..211...24M}, and \citealt{Reinhold&Laurent2015}).  The method of determining stellar ages from rotation rates is termed Gyrochronology (\citealt{Barnes2003ApJ...586..464B}). Asteroseismology, which can be 
applied to measure stellar rotation, assists typically in accurately constraining ages of cool stars.

We investigated how rotation rates correlate with their ages for this ensemble.
We show the age dependence of the mean envelope rotation rates of our sample in figure \ref{fig:skumanich}.
We fit a power-law function of form $\langle f_{\rm rot} \rangle ({\rm d^{-1}}) = \alpha (\tau/{\rm Gyr})^\beta$ to the age-rotation characteristics. Since both uncertainties in ages and rotations are non-negligible, we adopted an error-in-variables approach to fit the data. Specifically we used the orthogonal distance regression (ODR) algorithm available in the Python package \texttt{Scipy} which implements the FORTRAN routine \texttt{ODRPACK} to minimize the perpendicular distance between observed data points and the test function. This approach is superior in the present case since it accounts for uncertainties in both the ages inferred and rotation rates. ODR fits our inferences with  $\{\alpha = 0.291 \pm 0.018, \beta = -0.048 \pm 0.016 \}$. The algorithm moreover provides the covariance matrix whose eigenvalues and eigenvectors represent the squared-length and directions of the principal axes of $1\sigma$ error ellipse (or, 68\% confidence interval). For the $2-\sigma$ error ellipse (95\% confidence interval), the principal axes scale by $\sqrt{5.991}$ in magnitude.
Figure \ref{fig:skumanich} conveys that the stars in our sample do not show significant spin-down, decidedly weaker than that for cool stars.

$\delta$ Scuti stars are A-F type stars and their large radiative envelopes or the tiny convective cores are unlikely to create strong dynamo magnetic fields. A small fraction ($\sim$10\%) of hot stars are known to possess strong magnetic fields (\citealt{2004Natur.431..819B_braithwaite}), called fossil magnetic fields, which are remnants of magnetic field from the protostellar clouds and are usually dipolar in morphology. Such fields, which have been observed in a very few $\delta$ Scuti stars (\citealt{10.1093/mnrasl/slv130, 10.1093/mnrasl/slx023}), have the capacity to permeate the entire star and subsequently transfer angular momentum (AM) to deeper layers, resulting in an effective loss of AM (\citealt{10.1093/mnras/staa237_effects_of_fossil}).
$\beta$ Cas (\citealt{2020A&A...643A.110Z}) is the only known $\delta$ Scuti to show the presence of a dynamo magnetic field.
Consequently, magnetic braking is generally not expected to play an important role in spin-down (\citealt{10.1093/mnrasl/slv130, Ahlers_2019}) of these stars.

Internal gravity waves can also drive AM transport in stars with convective cores and radiative envelopes, including $\delta$ Scuti stars and B-type stars (\citealt{Fuller_2014, Rogers_2015}). These waves transport AM from the location at which they are excited (core-envelope interface) to the convectively stable envelope, where they undergo dissipation and deposit AM.  Due to its stochastic nature, this mechanism can alternately increase or reduce the envelope rotation rate (\citealt{Rogers_2015}). The dynamics are inherently two-dimensional and cannot be simulated in 1D stellar models such as the ones we consider here. Examining the roles of internal gravity waves in the spin-down of $\delta$ Scuti stars is beyond our current scope.

\begin{figure}
    \centering
    \includegraphics[width=0.6\linewidth]{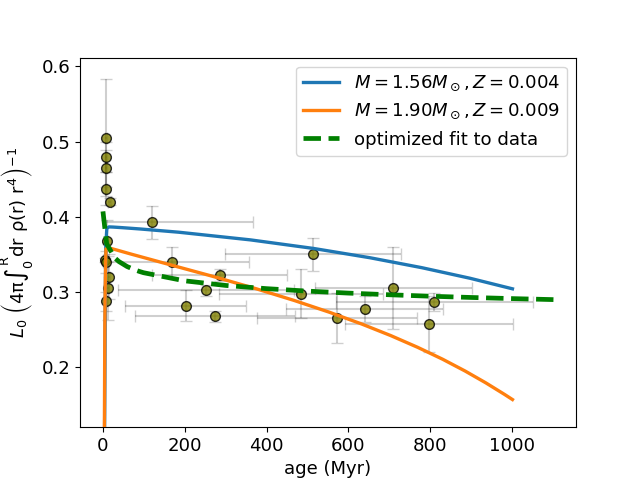}
    \caption{Age-dependent variation of the inverse moment of inertia ($I^{-1}(t)$) of two stars of different masses, scaled by arbitrarily assumed initial AM $L_0 = $ 0.4 and 0.6 $ M_\odot R_\odot^2 d^{-1}$ for stars of masses $1.56M_\odot$ and $1.90M_\odot$ respectively. In early phases, stars pass through the pre-main sequence phase of rapid contraction, which reduces their moment of inertia and causes a sharp rise in the rotation rate. During the main sequence this quantity gradually drops as a consequence of the rising momentum of inertia resulting from the radial expansion of these stars. In the background, we plot the rotation rates of the observed stars (figure \ref{fig:skumanich})  and  the best fit model for direct comparison.}
    \label{fig:rotMesa}
\end{figure}

Below, we estimate the extent of spin-down in typical $\delta$ Scuti stars in purely hydrodynamical context. Over the evolution through the main sequence, stars swell in size, and their momenta of inertia ($I(\tau) = 4\pi \int_0^{R_\star(\tau)} \rho(r,\tau) r^4 ~dr$) systematically increases. Assuming rigid body rotation and conservation of AM ($L_0$) throughout the evolution, the rotation rate ($\Omega(\tau) = L_0 / I(\tau)$) is expected to decay with stellar age, commensurate with $I^{-1}(\tau)$. Below, in figure \ref{fig:rotMesa}, we show the inverse moment of inertia for two different stars, scaled by arbitrary initial AM ($L_0$), as a function of stellar age. This explanation is able to describe the observed  weak-to-no spin-down, in the absence of explicit AM loss mechanisms.

\section{Conclusion}

We present a catalog of 5381 TESS $\delta$ Scuti stars of which 300 stars show pulsations with equally spaced frequencies.
As \cite{Bedding2020} stated for their sample and as we demonstrate in Figure \ref{fig:hrd1}, the dominant fraction of our population appears to lie closer to the ZAMS line, implying that they are young. However, non-perurbative simulations in ellipsoidal geometry (\citealt{Lignières2006, Reese2006, Reese_2008, Reese2009}) indicate that the island-mode pulsations of $\delta$ Scuti stars also exhibit uniform frequency spacings. Therefore, regular frequency patterns are likely present in evolved stars as well; however, the impact of mixed modes, avoided crossings and rapid rotation (\citealt{cepher}) could hinder their detection. Hence, in addition to the commonly assumed prevalence of regularly spaced oscillations in young stars, this sample of 300 stars may even contain a few evolved stars.

We exploited the available photometric, spectroscopic, and derived asteroseismic properties for this limited sample of 300 stars to place model-dependent constraints on their masses, metallicities, and ages.  The uncertainties in ages for older stars are considerably larger. We used a grid of non-rotating stellar evolutionary models (\citealt{murphy2023grid}) to derive the best-fit parameters for these stars (\citealt{structure_ML}), finding most of them to be young and show sub-solar metallicities ($Z<0.020$) along with a few exceptions to this. However, the absence of rotation in stellar models can lead to inaccurate inferences. For example, rotation can enhance mixing processes inside stars, resulting in faster evolution. Therefore, older stars causing the bimodal distribution of ages in Figure \ref{fig:modelfit} may be due to non-rotating models systematically indicating larger ages for some stars. Conversely, figure \ref{fig:hrd1} shows some of the 300 stars lying away from the ZAMS line, partially supporting their evolved nature. Without a grid of rotating models at our disposal, we cannot robustly estimate the reliability of the inferred ages. Thus, our current results should be viewed as a first effort at confronting the observations with non-rotating stellar models.

In the oscillation spectra of 24 stars among the 300, we detected rotational splittings in the dipole modes and inferred their mean rotation rates in a model-independent manner. Although the rotation rates displayed a correlation with age, the large uncertainties in the latter leave us with statistically insignificant rotational decay, at least in comparison with spin-down rates seen in the case of cool stars. The hydrodynamical evolution in the moment of inertia causes a  natural decay in the rotation rates - and is consistent with the measurements. Thus, we were unable to constrain the extents of the effects of alternate processes, such as gravity modes or magnetism. Nevertheless, direct evidence of both fossil (\citealt{10.1093/mnrasl/slv130, 10.1093/mnrasl/slx023, ds_4_mgntic}) and dynamo (\citealt{2020A&A...643A.110Z}) magnetic fields, as well as indirect effects such as spots (\citealt{Balona2019AFspot}) and flares (\citealt{Balona2012AFFlare}) have been detected in some $\delta$ Scuti stars. However, the magnetic fields in these stars are relatively weak (\citealt{B_dlt_str}). Such field may  induce magnetic braking, slowing down the stars, but whether this is a significant contribution remains to be determined.

In future, we plan to generate a grid of rotating stellar models and use it to infer more reliable ages for $\delta$ Scuti stars. This may provide a more definitive picture on the gyrochronology of $\delta$ Scuti stars. Our present sample is limited to slowly rotating stars in the near-ZAMS phase and hence does not capture the complete landscape of rotation rates or spin evolution of the full $\delta$ Scuti class. While slowly rotating stars make it easier to discern rotational splitting, near-ZAMS $\delta$ Scuti stars allow for seismologically constraining their ages through the regular oscillation patterns they exhibit. Given the paucity of discernible rotational splittings in $\delta$ Scuti pulsations, we plan on using spectral-line broadening in order to estimate the rotation rates of a large number of $\delta$ Scutis. \citet{cepher} have shown that the spectroscopically measured line-broadening parameter (\texttt{vbroad}), available in Gaia DR3, serves as a reliable measure of stellar rotation. This, in conjunction with improved stellar-age inferences, may enable the study of larger samples of $\delta$ Scuti stars.

\section*{Supplementary Information}
\restartappendixnumbering
\renewcommand{\thefigure}{S\arabic{figure}}
\setcounter{table}{0}
\renewcommand{\thetable}{S\arabic{table}}
\setcounter{figure}{0}
\subsection*{Appedix A: Removing External periodicity}\label{Appendix:Methods external periodicity}
We first eliminated externally arising sources of periodicity from binary and multiple-star systems from our sample by cross-matching with the Washington Double Star (WDS) (\citealt{wds_2}) catalogue. This step was crucial in preventing the introduction of additional frequencies or regular patterns due to extrinsic brightness variations and also to avoid misinterpreting modes.

Further, to decrease the possibility of false positives, 
we cross-matched our sample with the SIMBAD (\citealt{simbad}) database for the object types mentioned below and subsequently discarded them:
Ellipsoidal Variable, Eclipsing Binary, Spectroscopic Binary, BY Dra Variable, Symbiotic Star, X-ray Binary, Low Mass X-ray Binary, High Mass X-ray Binary, Cataclysmic Binary, Classical Nova, Double or Multiple Star, RR Lyrae Variable, Cepheid Variable, Classical Cepheid Variable, beta Cep Variable, Evolved Supergiant, Red Supergiant, Yellow Supergiant, Blue Supergiant, Wolf-Rayet, Neutron Star, Pulsar, Red Giant Branch star, Hot Subdwarf, Carbon Star, S Star, Long-Period Variable, Mira Variable, OH/IR Star, Post-AGB Star, RV Tauri Variable, Planetary Nebula, White Dwarf, Low-mass Star, Brown Dwarf, Extra-solar Planet, Eruptive Variable, Rotating Variable, Irregular Variable, High Proper Motion Star, Cluster of Stars, Globular Cluster, Open Cluster, Association of Stars, Stellar Stream, Moving Group, Type II Cepheid Variable.

Finally, we calculated the Renormalized Unit Weight Error (RUWE) from GAIA-dr2 (\citealt{gaia_release, gaia}), and removed the stars with RUWE greater than 2.0, as it may indicate the presence of binary and multiple-star systems.
It is derived by adjusting the Unit Weight Error (RUWE), a statistical measure of a star’s goodness of fit to GAIA's five-parameter astrometric model, encompassing positions, parallaxes, and proper motions. A RUWE of 1.0 ideally characterizes a well-behaved single star. However, in practice, RUWE values often deviate due to extreme magnitudes and colors. To address this, RUWE is re-normalized through division by a reference RUWE value of a carefully chosen well-behaved star, whose magnitude and colour are stable. Stars with RUWE greater than 2.0 may indicate the presence of a binary or multiple-star system (\citealt{Evans_2018}). Thus, we removed stars with RUWE values greater than 2.0 from our sample. 

\subsection*{Appendix B: Additional Figures}
\begin{figure}
    \centering
    \includegraphics[width=0.4\textwidth]{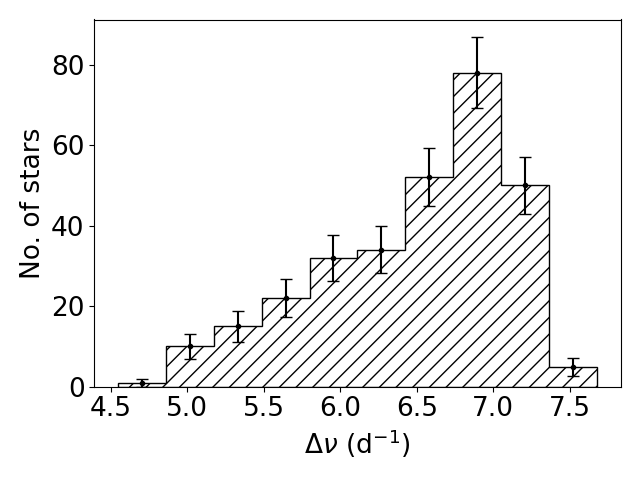}
    \caption{Distribution of $\Delta \nu$ of all the 300 $\delta$ Scuti stars exhibiting regular oscillation patterns.}
    \label{fig:dnu_histogram}
\end{figure}

\begin{figure}
    \centering
    \includegraphics[width=0.6\textwidth]{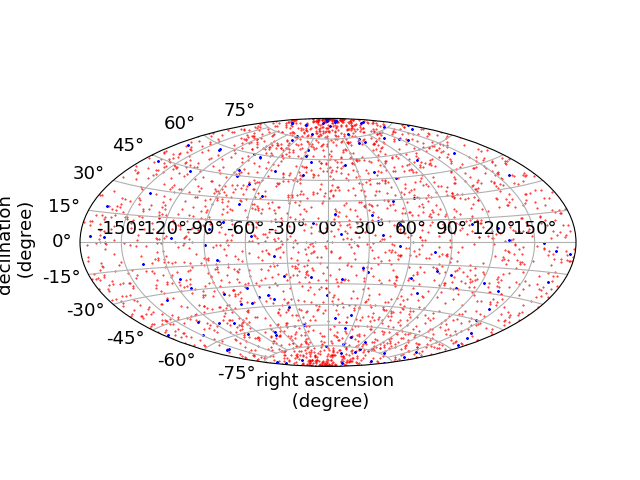}
    \caption{Sky map of $\delta$ Scuti stars from the first 63 sectors of TESS in right ascension and declination. The red dots represent all the $\delta$ Scuti stars and the blue dots represent regular pulsators.}
    \label{fig:skymap1}
\end{figure}

Figure \ref{fig:dnu_histogram} shows the distribution of $\Delta \nu$ values for all the 300 regular $\delta$ Scuti Stars. The median value of $\Delta \nu$ for these stars is 6.66$d^{-1}$.

\begin{figure*}
    \centering
    \includegraphics[width=0.8\textwidth]{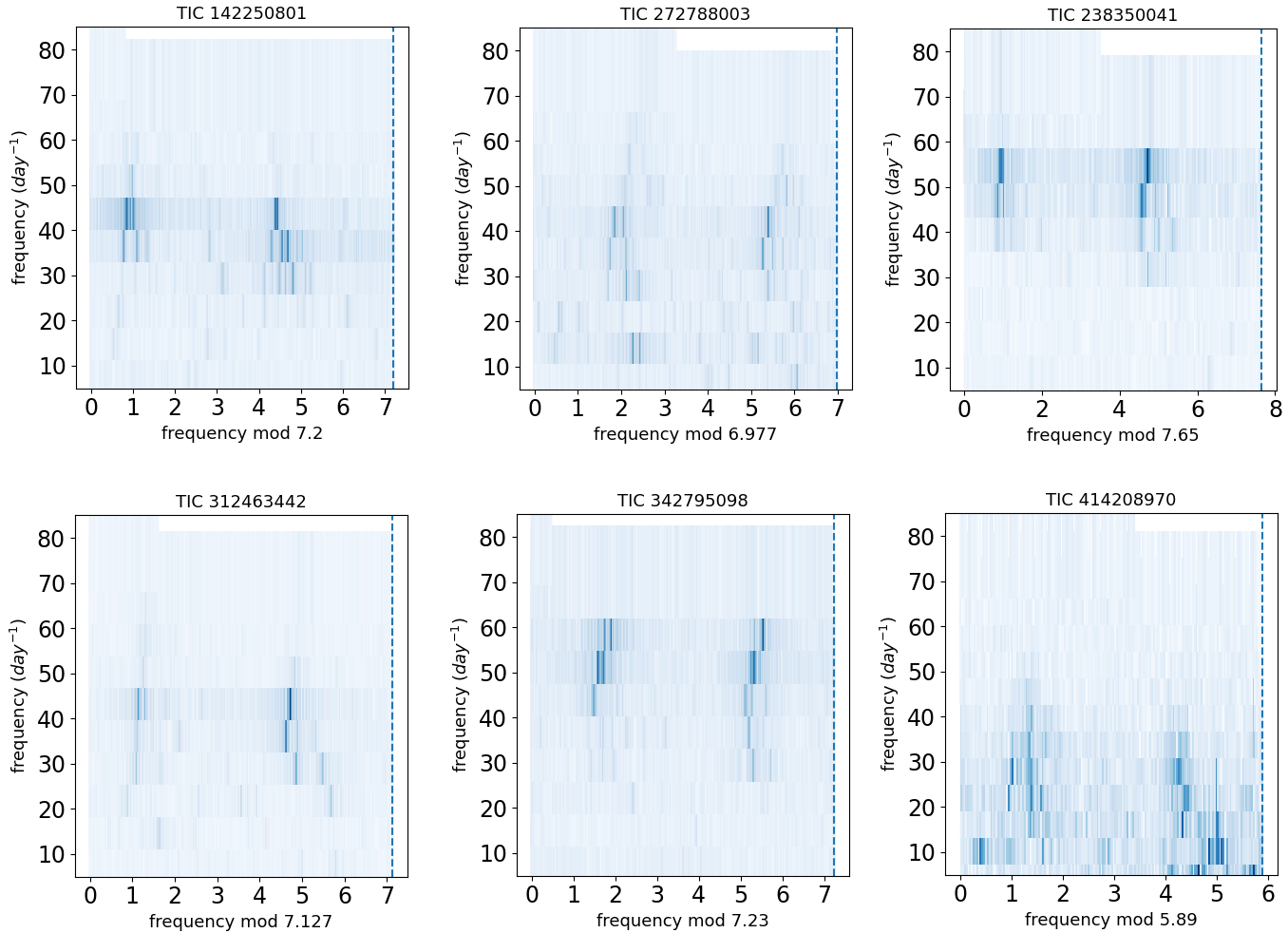}
\caption{\'echelle diagrams of $\delta$ Scuti stars exhibiting regular frequency separations in their oscillation spectra.}
\label{fig:regular_stars_echelle_diagram}
\end{figure*}

In figure \ref{fig:regular_stars_echelle_diagram}, we present several examples of regular $\delta$ Scuti stars with distinct vertical ridges. The ridge on the right corresponds to the $\ell=0$ mode and the one on the left to the $\ell$=1 mode.

\begin{figure}
    \centering
    \includegraphics[width=0.56\textwidth]{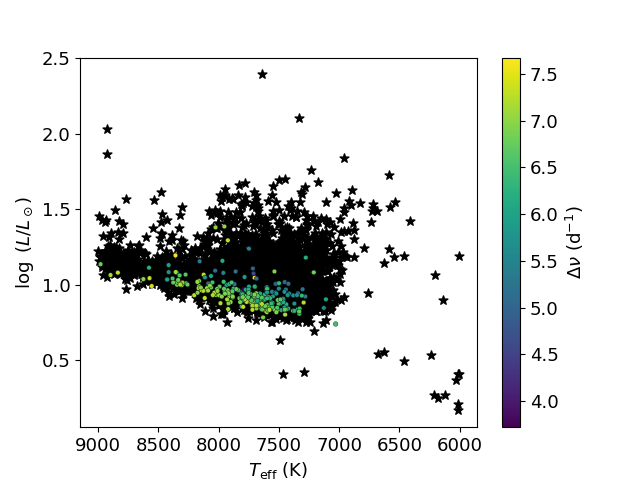}
    \caption{HR diagram of 5381 $\delta$ Scuti stars (black dots) and 300 regular pulsators overplotted with color-coding according to their $\Delta \nu$ values.}
    \label{fig:hrd1}
\end{figure}

Figure \ref{fig:hrd1} indicates that most regularly pulsating $\delta$ Scuti stars lie near the main sequence and stretch into the instability strip.

\begin{figure*}
    \centering
    \includegraphics[width=0.68\textwidth]{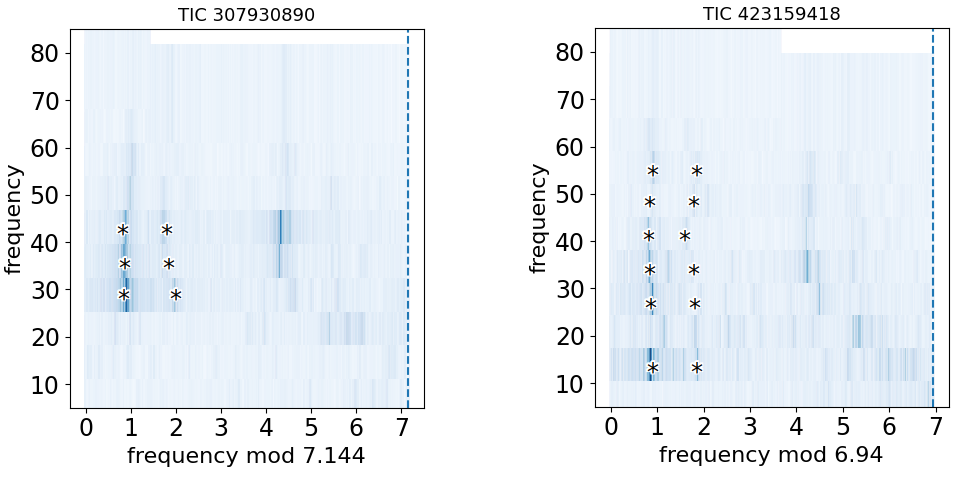}
    \caption{\'echelle diagrams illustrating rotational splitting in dipole modes}
    \label{fig:rotational_splitting_good}
\end{figure*}

Figure \ref{fig:rotational_splitting_good} illustrates two examples from the sample, in which rotational splitting may be seen across all radial orders. 

\begin{figure}
    \centering
    \includegraphics[width=0.56\textwidth]{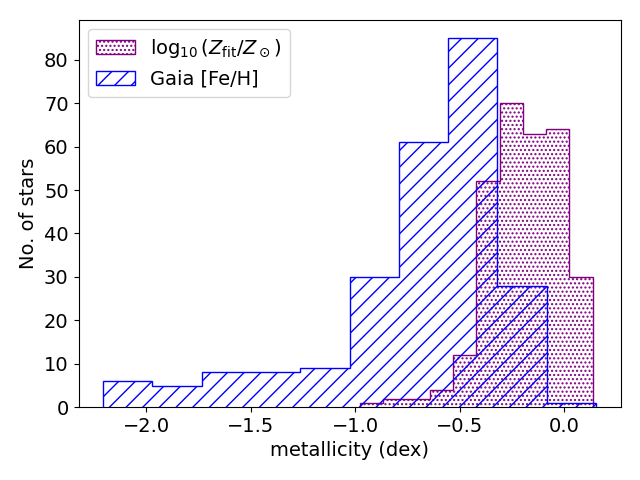}
    \caption{
    Comparison between distributions of the metallicity calculated from the fitted model using ${\rm [Fe/H]} = \log_{10}(Z/Z_\odot)$ and GAIA DR3 $\tt{mh\_gspphot}$. It appears that our inferences overestimate the metalicities of the examined stars. The reason is not straightforward to guess, however we presume while the Gaia provided [Fe/H] is spectroscopically measured from the atmospheric Fe abundance, our inference represents the initial global metallicity $Z_i$. Regardless this discrepancy, both ours and Gaia data suggest that most of the values remain below the solar metalicity $Z_\odot$ (or [Fe/H]=0).
    }
    \label{fig:Zcompare}
\end{figure}

We were able to obtain GAIA DR3 metallicities (described as $\tt{mh\_gspphot}$) for some stars in our sample. We compared their statistics against the distributions of our inferred metallicities, shown in figure \ref{fig:Zcompare}. The former refers to surface metallicity, while the latter points to global metallicity. Surface and global metallicities of stars are expected to remain identical due to mixing processes, provided no significant mechanisms hinder this homogenization.

\begin{figure*}
    \centering 
    \includegraphics[width=\textwidth]{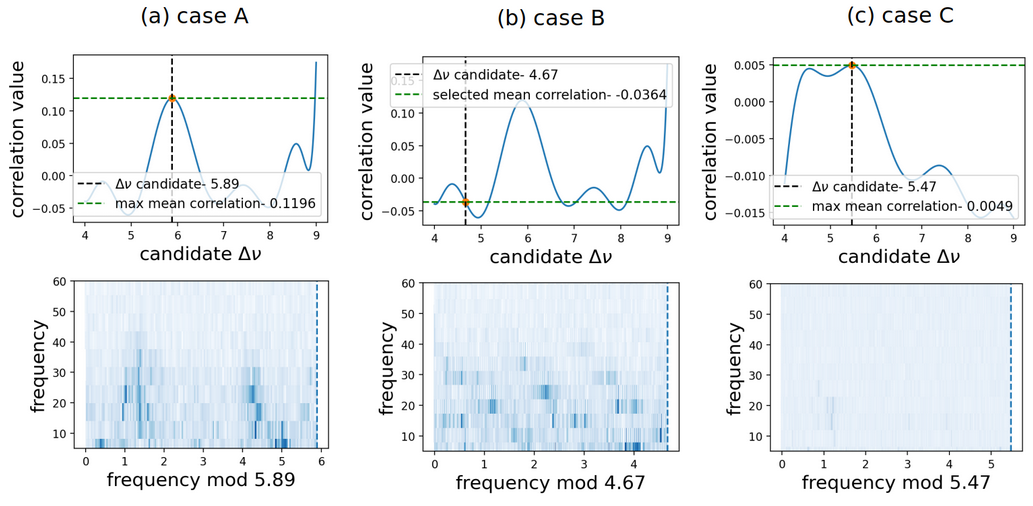}
    \caption{Case A displays the correlation graph of a regular star with the chosen candidate $\Delta \nu$ corresponding to the global maximum and the resulting \'echelle diagram. \\
    Case B displays the correlation graph of a regular star with the selected candidate $\Delta \nu$ corresponding to a lower correlation value and the resulting \'echelle diagram. \\
    Case C displays the correlation graph of an irregularly pulsating star with the selected candidate $\Delta \nu$ corresponding to the global maximum and the resulting \'echelle diagram.}
    \label{fig:corr_graph comparison}
\end{figure*}

The figure \ref{fig:corr_graph comparison} shows 3 cases of the auto-correlation algorithm introduced in section 2.3 to sort the stars in ascending order of their spectral regularity and simultaneously determine their large frequency separations ($\Delta\nu$). \\

\begin{figure*}
    \centering
    \includegraphics[width=\textwidth]{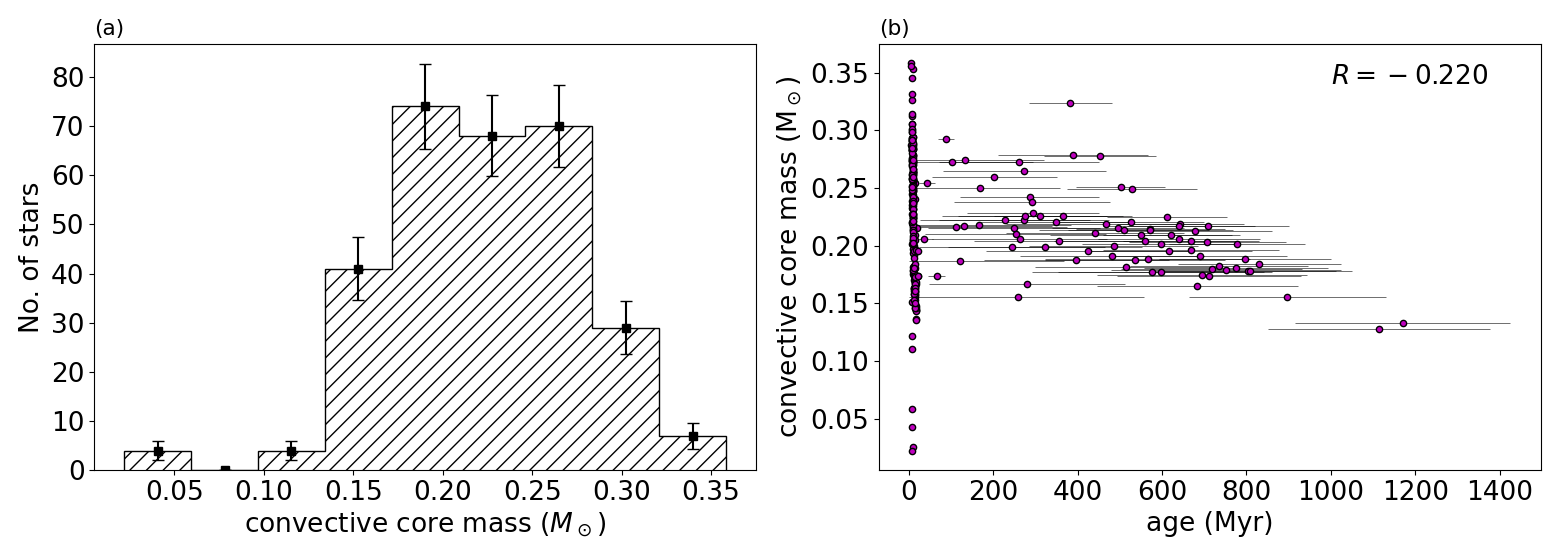}
    \caption{(a) Distribution of convective core masses of the regularly oscillating $\delta$ Scuti stars. (b) Evolution of convective core masses with stellar ages,  showing a declining correlation of Pearson $R = -0.220$. As a consequence of the depletion of core Hydrogen during the main sequence, the mass of the convective core tends to decrease with evolution. The correlation $R$ turns out small due to the scattered presence of very young stars, ignoring which $R$ coefficient raises to $-0.583$.}
    \label{fig:colvMcore}
\end{figure*}

The convective cores of $\delta$ Scuti stars play an important role in their evolution, as they generate the entire energy necessary to counteract the gravitational collapse. We evolved all 300 regularly pulsating stars using their best fit parameters and determined the mass of their convective cores. We show its distribution and age-dependence in figure \ref{fig:colvMcore}.

\begin{figure*}
    \centering
    \includegraphics[width=\textwidth]{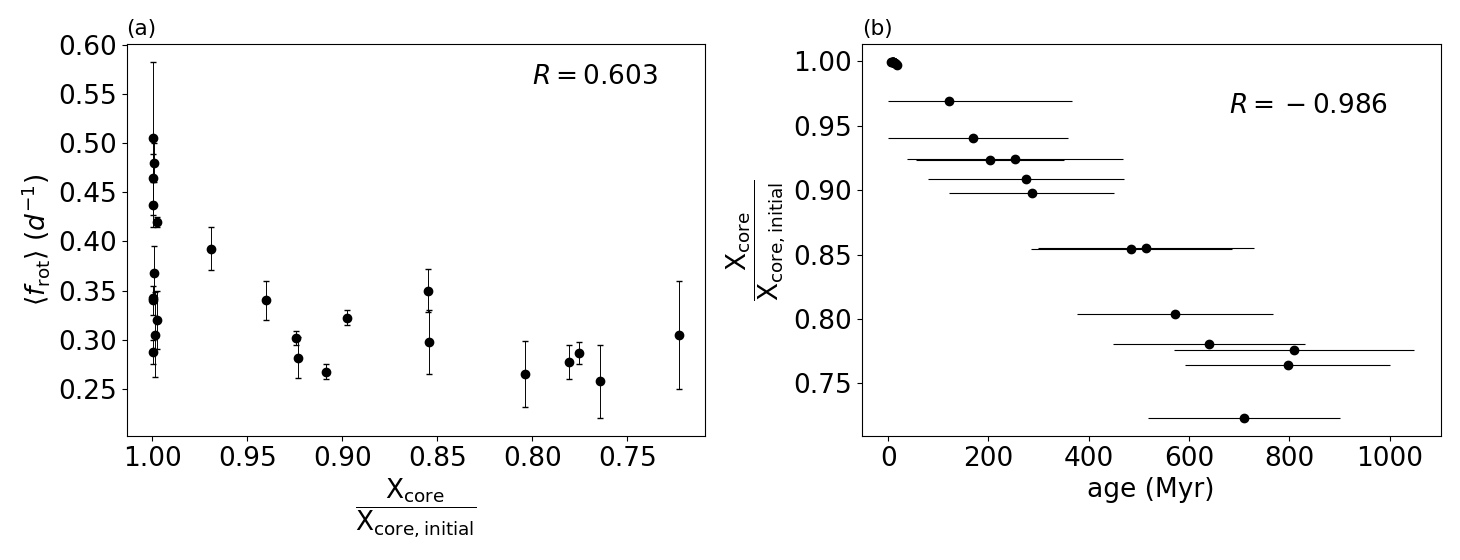}
    \caption{(a) Average envelope rotation as a function of core Hydrogen content $X_c$ (normalized by the initial value $X_{c,i}$) of the 24 stars that show rotational splitting.  Normalized core Hydrogen content serves as the most reliable representation of the phase of stellar evolution,
    The correlation coefficient R of $0.603$ between two quantities implies that the dependence is likely not spurious. The correlation is diminished by the cluster of the youngest stars at the left. (b) Evolution of core Hydrogen abundances of the same stars with their ages. The strong anti-correlation with $R = -0.986$ delineates the tight dependence of core Hydrogen abundance over stellar ages, the former linearly declining while the latter advances.}
    \label{fig:coreHrotation}
\end{figure*}

It is more informative to study the spin evolution of stars with their evolutionary states. Core Hydrogen content is one of the direct indicators of stellar evolution as it provides the fuel budget necessary to support various dynamics of the stars. We evolved the 24 stars, that show rotational splittings, using their best fit structures and obtained their core Hydrogen abundances. We show in figure \ref{fig:coreHrotation} the dependence of average stellar rotation with the core Hydrogen content and the evolution of the latter with the stellar ages.

\section*{Supplementary Resources:}
The following supplementary files are available at
Zenodo: doi:\href{https://doi.org/10.5281/zenodo.15270349}{10.5281/zenodo.15270349}.

\begin{enumerate}[nolistsep]
    \item csv file (\texttt{Supp1\_TIC\_Ids\_5381\_star.csv}) with TIC IDs of 5381 $\delta$ Scuti stars identified from the first 63 sectors of TESS
    \item csv file (\texttt{Supp2\_regular\_300\_stars\_parameter.csv}) containing various properties of the 300 $\delta$ Scuti stars that exhibit regular oscillation. \\
        There are 17 fields in this file, which from left to right contain: \\
        TIC IDs, HD IDs (-1 when not available), Gaia G-magnitude Gmag (nan when not available), error in Gmag (nan when not available), Gaia Bp-Rp color (nan when not available), $\Delta\nu$ (${\rm d}^{-1}$), error in $\Delta\nu$ (${\rm d}^{-1}$), best-fit $M$ (in $M_\odot$), lower uncertainty in $M$ (in $M_\odot$), upper uncertainty in $M$ (in $M_\odot$), best-fit $Z$, lower uncertainty in $Z$, upper uncertainty in $Z$, best-fit age (in Myr), lower uncertainty in age (in Myr), upper uncertainty in age (in Myr), $\chi^2_{\rm min}$ of the best-fit model.
    \item pdf file (\texttt{Supp3\_echlle\_spectra\_with\_model\_modes.pdf}) with \'echelle diagrams and power spectra of 300 $\delta$ Scuti stars with regular pulsation:\\
        The best-fit model's properties are indicated on each spectrum. The radial and dipole mode frequencies of the best-fit models are also shown over the \'echelle diagrams with circles and triangle symbols respectively.
    \item csv file (\texttt{Supp4\_rot\_rates\_24\_star.csv}) containing Ids of 24 $\delta$ Scuti stars with rotational splittings along with their rotation rates. \\
        This file contains 6 fields, which from left to right indicate: \\
        TIC Id, large frequency separation $\Delta\nu$ (${\rm d}^{-1}$), mean rotation rate (${\rm d}^{-1}$), error in mean rotation (${\rm d}^{-1}$), error in $\Delta\nu$ (${\rm d}^{-1}$), and  core hydrogen content for the star.
    \item csv file (\texttt{Supp5\_modes\_info\_rot\_24\_star.csv}) providing details of rotational splittings seen in the 24 stars.
        This file contains 10 fields, which from left to right indicate: \\
        TIC Id, radial order $n$, Ledoux constant for radial order $n$, frequency of $(n, \ell=1, m=-1)$ (${\rm d}^{-1}$), frequency of $(n, \ell=1, m=0)$ (${\rm d}^{-1}$), frequency of $(n, \ell=1, m=+1)$ (${\rm d}^{-1}$), rotation rate $f_{\rm rot}$ (${\rm d}^{-1}$) derived for radial order $n$, median rotation rate (${\rm d}^{-1}$), uncertainty in rotation rate (${\rm d}^{-1}$), core hydrogen content for the star.
    \item pdf file (\texttt{Supp6\_rot\_split\_zoomed.pdf}) showing zoomed-in view of rotationally split frequency multiplets of the 24 stars.
\end{enumerate}

\section*{Softwares used:}
\begin{enumerate}[nolistsep]
    \item \textsc{Numpy} (\citealt{harris2020array})
    \item\textsc{Scipy} (\citealt{2020SciPy-NMeth})
    \item \textsc{Lightkurve} (\citealt{2018ascl.soft12013L})
    \item \textsc{scikit-learn} (\citealt{scikit-learn})
    \item \textsc{matplotlib} (\citealt{Hunter:2007})
    \item \textsc{multiprocessing} (\citealt{multiprocessing_mckerns2012building})
    \item \textsc{pandas} (\citealt{mckinney2010data})
    \item \textsc{astroquery} (\citealt{2019AJ....157...98G})
\end{enumerate}

\section*{acknowledgments}

We acknowledge several useful discussions with Prof. Jørgen Christensen-Dalsgaard, Dr. Simon J. Murphy and Prof. Timothy R. Bedding. We also thank Prof. Michel Rieutord for their useful comments. We acknowledge support from the DAE, Government of India (Grant No.: RTI 4002). This analysis has made use of the data collected from the TESS mission, obtained from the MAST data archive at the Space Telescope Science Institute (STScI). Funding for the TESS mission is provided by NASA's Science Mission Directorate. This study also made use of the data from the SIMBAD database, operated at CDS, Strasbourg, France, the VizieR catalogue access tool, CDS, Strasbourg, France and the Washington Double Star Catalogue maintained at the U.S. Naval Observatory. Furthermore, our work made use of data from the European Space Agency (ESA) space mission Gaia (\href{https://www.cosmos.esa.int/gaia}{https://www.cosmos.esa.int/gaia}). Gaia data are being processed by the Gaia Data Processing and Analysis Consortium (DPAC). This research was supported in part by a generous donation from the Murty Trust, aimed at enabling advances in astrophysics through the use of machine learning. Murty Trust, an initiative of the Murty Foundation, is a not-for-profit organisation dedicated to the preservation and celebration of culture, science, and knowledge systems born out of India. The Murty Trust is headed by Mrs. Sudha Murty and Mr. Rohan Murty. We are grateful to the referee for their constructive comments.

\end{document}